# MOBILE BLOCKCHAIN DECENTRALIZED APPLICATIONS (DAPPS): A CASE STUDY OF IPTM BLOCKCHAIN CERTIFICATE VERIFICATION SYSTEM

Ts. Dr Mohd Anuar Mat Isa[1], iExploTech

Dr. Muzaffar Hamzah, Universiti Malaysia Sabah (UMS)

Daimler Benz Alebaba, iExploTech

## Introduction

A variety of mobile devices and applications have spread the usability of blockchain solutions to over 5.27 billion unique mobile phone users (Kepios, 2021). The rising of Bitcoin price up to USD 50,000 in March 2021 has made many blockchain mobile wallets and smart contracts DApps popular for current and future investment of cryptocurrency and digital-asset managements. To understand the trend, this chapter will present the design and implementation of mobile blockchain DApps using Android Studio together with Ethereum's smart contract as the digital-asset management tool. Java Android and Ethereum's Web3-Java APIs will be demonstrated as a practical deployment of the mobile DApps. The logic and decision-making of the mobile DApps will be demonstrated and coded as a smart contract. The source codes of the mobile DApps and smart-contract were published in Github[2] as open-source codes for those who are interested to build and run the project.

To dive into the state of Blockchain technology, this chapter presents a case study to assess the usability and practicality of the blockchain DApps. The case study will demonstrate how to verify academic credentials or certificates using blockchain technology. This certificate verification DApps was designed using Ethereum's smart contract and it was programmed using a Solidity programming language. This smart contract namely, *"IPTM_BlockchainCertificate.sol"*[3] is downloadable from GitHub repository under an open-source license. A client-side for checking the validity of certificates was implemented using Android Studio namely, *"IPTM_Certificate"*[4] which is also available for download in the GitHub repository under an open-source license. To facilitate the entire ecosystem of the certificate verification DApps, iExploTech has shared two developer tools and a system admin manual that is accessible to the public, namely *"IPTM_Certificate_DB_Tool.jar"*, *"IPTM_Certificate_Generator_Tool.jar"* and *"IPTM_Admin_Manual.pdf"*. These tools with the admin manual will assist the IT department in the university to set up and deploy the entire system. Figure 1 illustrates components and interactions of the IPTM certificate verification system.

---

[1] Corresponding author
[2] https://github.com/iexplotech/SmartContract
[3] https://github.com/iexplotech/SmartContract/blob/master/IPTM_BlockchainCertificate.sol
[4] https://github.com/iexplotech/IPTM-Blockchain-Certificate-Verification-System-Android

Before this chapter focus to the case study, a reader should grasp the state of the art that makes blockchain technology emerged. Thus, this chapter is organized into eight (8) sections for the reader to easy to follow and comprehend the knowledge. The introduction section provides an overview of this chapter's contents. The second section, Related Work revisits the blockchain and distributed ledger technology (DLT) as well as a debut history of Ethereum's smart contract by Vitalik Buterin. The third section, IPTM Background reveals the motivation and background of the Institusi Pendidikan Tinggi Malaysia (IPTM) blockchain network. Forth section, IPTM Certificate Design enlightens the design of the entire IPTM certificate verification system. The fifth section, IPTM Certificate Experimental Setup elaborates in detail on the scope and experiment setup for measuring the performance of the IPTM certificate verification system at the minimal cost for deployment. The sixth section, Experiment Result presents the experimental results that centering on the stress of the IPTM's blockchain sealer node to provide read and write certificate requests. The seventh section, Discussion reflects the outcome of experimental results. The final section, Conclusion concludes the potential of blockchain implementation for the credential and certification verification system.

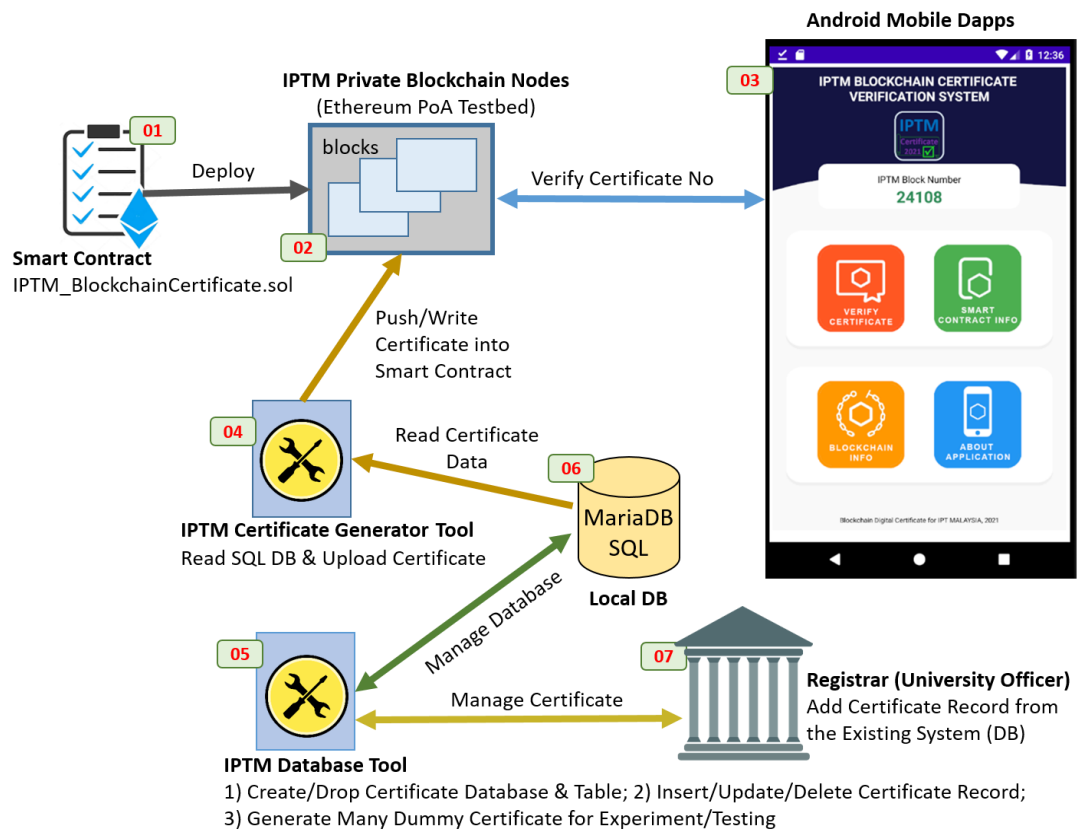

Figure 1. An overview of the IPTM certificate verification infrastructure provided by iExploTech for universities in Malaysia

## Related Work

The first blockchain protocol and infrastructure was implemented as Bitcoin (Nakamoto, 2009). Bitcoin relies on elliptic curve cryptography (ECC) security (secp256k1 curve), as a digital signature scheme for all Bitcoin transactions. Before a transaction or data write into blockchain as a legitimate transaction, these data must be encoded and hashed using Standard Hashing Algorithm 256 (SHA256). The owner of Bitcoin will use his private key to sign the SHA256 hash with a unique nonce for each transaction. Almost all transaction in Bitcoin is used for transfer Bitcoin to other accounts and also his account (split transaction).

After signing the transaction, the owner (using Bitcoin client) will broadcast his transaction to all reachable Bitcoin nodes using a peer-to-peer protocol. Before the other Bitcoin nodes approve (write) this transaction into the latest Bitcoin mining block, they will verify the signed transaction using a public key of the transaction owner as well as the unique nonce. The transaction will be parked or stored (mining) in the latest block if the transaction is successfully verified by the owner digital signature (condition 1) together with a sufficient amount of payment Bitcoin transaction fee (condition 2) as well as having enough Bitcoin amount for transfer (condition 3). From here one may observe that Bitcoin heavily relies on cryptography to enable the security and trustworthiness of any Bitcoin transaction.

Before Ethereum (Buterin, 2013) was introduced, Vitalik Buterin has discussed with his Bitcoin development team for a new blockchain feature. He suggested an upgrade to the Bitcoin network which could allows users to write a piece of code in the Bitcoin network. However, his suggestion was rejected a couple-time and he finally decided to move on by modifying Bitcoin client to implement his brilliant idea. That was the foremost inauguration of a smart contract implementation in the blockchain network. Comparing to the Bitcoin protocol, Ethereum protocol uses the same ECC's secp256k1 curve for the digital signature, but Ethereum has selected Keccak256 (not NIST-SHA3) (Yaga et al., 2019) as the hashing algorithm. To enable one to write a code in the Ethereum network, Gavin Wood has designed a specific programming language for one to write a smart contract, namely Solidity (Wood, 2014). He has assisted Vitalik Buterin to realize the idea of allowing everyone to write code in the blockchain using the Solidity language.

The flashback in the two earlier paragraphs has given insight into the founder's determination to make today's blockchain technology. This paragraph will further explore existing works that are related to the implementation of smart contracts for credential or certificate verification. There are quite a number of research publications discuss on certificate verification using blockchain technology, however, most publications discuss on a surface without a deep understanding and practical experience on the subject matter. Majority of published papers by these novice authors were presented based on what they have learned by reading some blockchain publications. The authors rephrase his finding into a new research publication with a basic idea or framework on how to incorporate a blockchain and smart contract technology. With an infancy level of knowledge and skill, they tried to solve issues they assumed to be a real-world problem by adding blockchain technology. They do not realize that mostly blockchain actual problems that required further research and investigation are available in the developer discussion chat-room and mailing list such as Bitcoin, Ethereum, and Hyperledger developer websites.

This paragraph listed the infancy quality publications of the blockchain certificate verification method. The authors (Cheng et al., 2018) show a framework and a snapshot of blockchain certificate verification by scanning a QR code and verified it through the website. However, there is no evidence, code or experiment result provided by the authors to mark the proposal is actually implemented. The authors (Poorni et al., 2019) presented almost similar to (Cheng et al., 2018), but just the diversity that (Poorni et al., 2019) implemented is the GUI using Java Swing's APIs for a desktop application to input certificate data. The authors (Nurhaeni et al., 2020; Vidal et al., 2020) suggest adopting blockchain as a method to provide tamper resistance for storing registry records and certificates. The authors (Gayathiri et al., 2020; Xie et al., 2020) show a framework and snapshot of blockchain certificate verification without using QR code, which is inferior compared to (Cheng et al., 2018). The authors (Zhao et al., 2020) propose to integrate Certificate Authorities (CA) with blockchain technology for storing certificates of public key infrastructure (PKI). All listed publications in this paragraph did not provide any significant practical work, coding, or experiment results to show their suggestion or proposal is a concrete statement rather than just barely hypothetical statement.

This paragraph listed significant publications for references. The authors (Liu & Guo, 2019) have deployed Hyperledger's chain-code for the certificate verification system. Based on the authors' experimental testbed, the performance of the transaction in Hyperledger is 180-250 transactions per second. There is no sample code provided as well as the authors have a misunderstanding that Hyperledger is a blockchain technology (grouping transactions at a certain time-frame), which actually Hyperledger is based on directed acyclic graph (DAG) technology. The authors (Bousaba & Anderson, 2019) have deployed Ethereum smart contract and execute the certificate's smart contract APIs through a web interface. The smart contract allows anyone with sufficient Ether to upload certification information whereby this information is encrypted by a user password. To read the certificate information, one must supply a correct password through the website. The proposed system does not allow public certificate verification because it is restricted by the user password to decrypt certificate data as well as there is no proper way to verify the uploaded certificate data is a genuine certificate issued by a university. The authors (Mishra et al., 2020) have deployed a certificate smart contract in Ethereum's Rinkeby testnet and have measure a gas cost to perform adding (write) a new certificate in the Rinkeby network. The average time is 16 seconds per certificate write (mined) under a proof-of-authority consensus (Rinkeby's PoA, 15 seconds per block). The authors (Gundgurti et al., 2020) demonstrated a digital certificate file was uploaded into Inter-Planetary File System (IPFS) using Remix IDE. The hash value of the digital certificate file was added into a smart contract as proof of a genuine digital document.

Referring to all aforementioned authors and literature, none of them has followed the standard practice of Ethereum's smart contract as suggested by professional practitioners and communities. The foremost principle is to present or expose the smart contract's code to public views (or limited view to all stakeholders) is compulsory to show the solidarity and trustworthiness of the digital contract. This will allow someone to verify the deployed binary blob in the blockchain which is identical to the source code (smart contract). The following statement quotes the ethic which demanding the author of a smart contract to expose the code for public views:

> *"When you expect a person to use a Smart Contract that you have written, either for money transfer or data storage, you are expecting them to digitally sign a transaction that could have significant consequences. Before you force that expectation on the users of your contracts, you should upload the identical source code to Etherscan so that you at least give the users an opportunity to audit the code to independently verify that the code that will execute actually does what you said it will do."* (Emmons, 2018)

## IPTM Background

Institusi Pendidikan Tinggi Malaysia (IPTM) Distributed Ledger Technology (DLT) Network or IPTM-DLT-Net is an open collaboration between researchers in Malaysia as shown in Figure 2. The core objective of the collaboration is to establish TestNet and MainNet that enable collaborative research, development and maintenance of DLT and Dapps applications among Malaysian universities. CyberSecurity Malaysia[5] and iExploTech[6] were appointed as the secretariat to coordinate the collaboration work. One of the major collaboration works by IPTM members is to establish a blockchain certificate verification system that is presented as a case study in this chapter.

---

[5] www.cybersecurity.my
[6] www.iexplotech.com

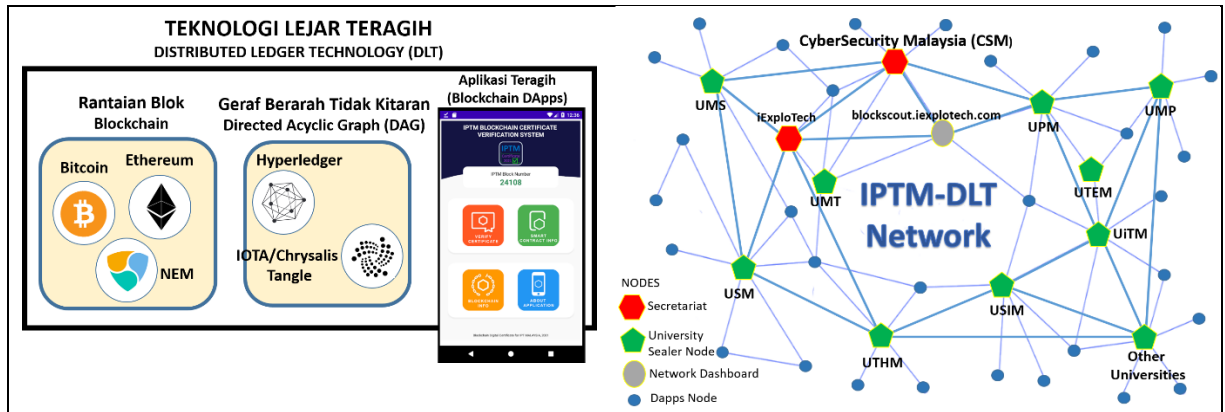

Figure 2. The collaboration between university's researchers in Malaysia to establish various DLT Networks

### IPTM's Certificate Design

Referring to Figure 1, there are seven (7) components were designed, developed and integrated by iExploTech. The most important component is the smart contract because its controls the logic and decision-making of the certificate verification system. Figure 3 shows the most frequent function calls executed for read and write certificate operations, namely *readCertificatePublic()* and *addCertificate()*. The *addCertificate()* function has a modifier that provides restriction to only the registrar address can execute this function. This restriction is very important to ensure that only the registrar from the university is permitted to upload genuine certificate information. The *readCertificatePublic()* function is a publicly accessible function for verifying the certificate number that mapped to a genuine certificate holder (owner). This function can be called by anyone using Android IPTM's DApps to verify the certificate owner. To explore further the seven (7) components as well as to set up the entire infrastructure, one may refer to the iExploTech's GitHub repository[7].

```
function readCertificatePublic(string memory _certNo) public view returns (
    string memory CertNo, string memory Name, string memory Programme, string memory ConvoDate) {

    // if studentId is empty, CertNo not exist
    if(strCompare(mapCert[_certNo].studentId , "") == 0)   // 0 is equal
        return ("", "", "", "");
    else
        return (_certNo, mapCert[_certNo].name, mapCert[_certNo].programme, mapCert[_certNo].convoDate);
}

function addCertificate(string memory _certNo, string memory _name, string memory _ic,
    string memory _studentId, string memory _programme, string memory _convoDate,
    string memory _semesterFinish) public onlyRegistrar returns (bool Status) {

    // Check CertNo existent, true if CertNo existed
    if(isValidCertificate(_certNo) == true) {
        return (false);  // may use revert()
    }

    mapCert[_certNo].name = _name;
    mapCert[_certNo].ic = _ic;
    mapCert[_certNo].studentId = _studentId;
```

Figure 3. Two important functions in the *IPTM_BlockchainCertificate.sol*[8] smart contract that frequently accessed by a registrar and end-user

---

[7] https://github.com/iexplotech

[8] https://github.com/iexplotech/SmartContract/blob/master/IPTM_BlockchainCertificate.sol

## IPTM Certificate Experimental Setup

This section describes the scope and experiment setup for measuring the performance of the IPTM certificate verification system at the minimal cost for deployment by universities in Malaysia. The scope of the experiment is to measure the performance of the IPTM sealer node for read and write certificate operations in the IPTM-DLT Network. Table 1 shows the testbed configuration for the experiment.

Table 1. Experimental testbed configuration

| Configuration | Descriptions |
|---|---|
| IPTM Blockchain Network (single version) | Blockchain client: Geth; private Ethereum PoA (Clique); 1 block every 5 seconds; 27,507,108 block gas limit (greater than public Ethereum mainnet), chain id: 496, pre-allocated ether at address 0x80ce17271ffa4a7f66e2cbf3561a6946587f470d (1 million ether) |
| IPTM Sealer Node (Geth) – iCore cloud virtual machine | 1 vCPU, 2 GB RAM, 50 GB RAID 10 HDD, 100 Mbps network bandwidth, one public IPv4, Debian 10 (Buster) x64 OS, iCore VM hosting[9] RM 30 monthly rental (~USD 7.28); cloud hosting location: Cyberjaya, Malaysia; Performance measurement tools: *iExploTech_Performance_Measurement_Server.jar* |
| Testing machine (send read & write requests to Sealer Node) | Intel i7-7700 CPU; 4 cores (8 threads); 12 GB RAM; Debian 10 (Buster) x64 OS; Telekom Malaysia fiber broadband: 100 Mbps network bandwidth; Physical testing machine location: Klang, Malaysia; Testing tools: *IPTM_Certificate_DB_Tool.jar* and *IPTM_Certificate_Generator_Tool.jar*; |

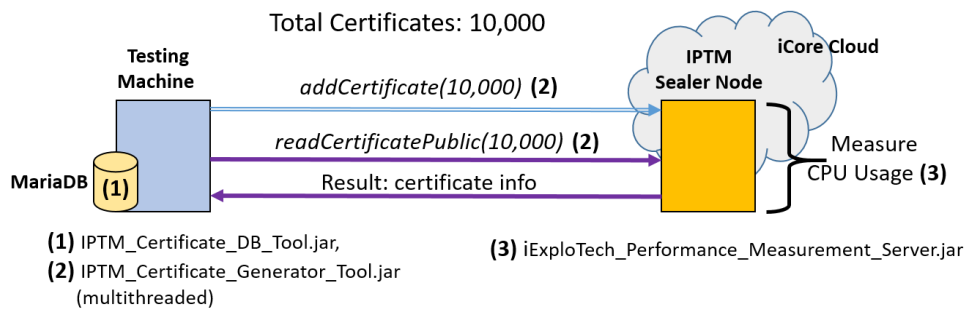

Figure 4. Experimental testbed setup

Referring to Figure 4, the IPTM's sealer node was tested to handle 10,000 certificate read and write operations. The *IPTM_Certificate_DB_Tool.jar* was deployed to generate 10,000 dummy certificate records in the MariaDB database in the Testing Machine. After that, the *IPTM_Certificate_Generator_Tool.jar* was deployed to read the 10,000 certificate records from the MariaDB database and then send all of them to the IPTM Sealer Node using the *addCertificate()* function (one by one certificate). After 10,000 certificates successfully write in the smart contract (blockchain), the *IPTM_Certificate_Generator_Tool.jar* was deployed once more to read the 10,000 certificates using the *readCertificatePublic()* function. During this read operation, the Testing Machine has executed this function by single and multithreaded requests. The *iExploTech_Performance_Measurement_Server.jar* in the IPTM Sealer Node was deployed to measure the CPU usage when the *addCertificate()* and *readCertificatePublic()* functions were tested.

---

[9] www.icore.com.my

Experiment Result

This section presents the performance of the *addCertificate()* and *readCertificatePublic()* functions to process the 10,000 dummy certificates. Before the execution of the *addCertificate()* function, the SQL query was successfully executed to read 10,000 records from the MariaDB database. The execution tooks less than 2 seconds to collect all certificate records. After that, the *addCertificate()* function was executed by 10,000 times to submitted all certificates using a single-threaded function call to the smart contract. It took 283.225847 seconds to submit all certificates. However, the IPTM Sealer Node only be able to write (confirm) the entire transactions by 622.296601 seconds. This happens because each Ethereum block has a transaction execution limit (block gas limit). Besides that, pending transactions will be written in the next block. On average, 80 certificate write transactions were confirmed per IPTM's block. Each IPTM's block is mined (found) at a fixed duration, every 5 seconds using Proof of Authority (PoA) consensus. Therefore, it can be roughly estimated that the performance of the *addCertificate()* function is 16 transactions per second[10] (TPS). To check the total available certificates in the smart contract, one may call the *getListCertificateStatus()* function to retrieve the certificate counter.

Table 2. The performance to read 10,000 certificates using parallel multithreaded requests

| Parallel Thread | Performance of Read 10,000 Certs (Seconds) | Testing Machine (Average CPU Usage %) | IPTM Sealer Node (Average CPU Usage %) | Average One Transaction Duration (Second) | Average Total Read Transaction Per Second |
| --- | --- | --- | --- | --- | --- |
| 0 | 0 | < 1 | < 1 | 0 | 0 |
| 1 | 168.791532 | 2.04 | 25.75 | 0.016731 | 59.24467822 |
| 2 | 73.71605 | 4.26 | 42.17 | 0.006605 | 135.6556679 |
| 3 | 55.337019 | 4.96 | 48.60 | 0.005346 | 180.7108547 |
| 4 | 44.096541 | 6.26 | 62.30 | 0.004267 | 226.7751568 |
| 5 | 37.741037 | 7.65 | 69.60 | 0.003654 | 264.9635727 |
| 6 | 38.427256 | 7.39 | 70.80 | 0.003595 | 260.2319562 |
| 7 | 35.015543 | 8.08 | 73.76 | 0.003394 | 285.5874604 |
| 8 | 29.957721 | 9.27 | 66.68 | 0.002854 | 333.803763 |
| 9 | 28.170607 | 10.59 | 73.37 | 0.002613 | 354.9799264 |
| 10 | 28.185154 | 9.87 | 79.13 | 0.00271 | 354.7967132 |
| 11 | 28.159086 | 10.43 | 73.39 | 0.002576 | 355.1251628 |

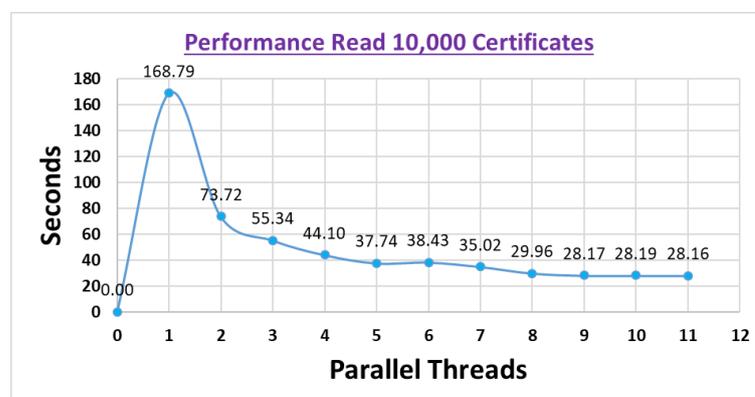

Figure 5. The graft of the performance to read 10,000 certificates using parallel multithreaded requests

---

[10] 80 transactions per block is divided by 5 seconds

After all certificates available in the smart contract, the experiment continued to perform read all 10,000 certificates using the *readCertificatePublic()* functions. The performance execution of this function is listed in Table 2 and Figure 5. The baseline of the performance measurement is the single thread execution which tooks 168 seconds to read the entire certificate with an average performance of 59 TPS. The CPU consumption of the Testing Machine ratio to the IPTM Sealer Node is 1:12, whereby a lot of computation performed by the blockchain node compared to the remote client that requested certificate verifications. The performance execution of multithreading reaches the maximum efficiency limit starting at nine (9) threads and above. The execution tooks 28 seconds to read the entire certificate with an average performance of 355 TPS. The CPU consumption of the Testing Machine ratio to the IPTM Sealer Node is 1:7. This reflects the parallel threads have reached the upper limit of optimum serving performance at the IPTM Sealer Node side. The average CPU usage of the IPTM Sealer Node was not reached beyond 90% because the geth client has a mechanism to restraint its execution from consuming the entire CPU resources. If anyone is interested to increase the transaction performance by double 355 TPS, an additional geth node with the identical hardware specification can be added to satisfies the requirement. If a new node joins into a blockchain network, it will automatically synchronize and replicate the entire blockchain by copying all blocks from other nodes using the P2P protocol. Therefore, it is easy to expand the blockchain system performance by adding multiple nodes that serving as a mirror database. This is the beauty of decentralized technology compared to a centralized system.

## Discussion

The Experimental Result section has shown the performance of the IPTM's Certificate Verification System using the most economic setup at a cost of about USD 7.26 monthly to maintain a blockchain node in the organization. For a small scale or an initial private blockchain deployment by a university in Malaysia, iExploTech recommends at least three (3) connected blockchain nodes that running as permanent nodes 24/7 to maintain the decentralized model. Other nodes can be added as a local mirror database machine using an old desktop computer.

Referring to the Related Work section, there are many existing works had implement certificate verification systems using blockchain technology. Yes, it is not a new topic as this chapter sharing knowledge with a reader. However, all of the previous works did not have a clear intention to spread the blockchain knowledge especially when they did not publish their smart contract for public views as well review it. Some of them share only the experimental result without allowing other researchers to replicate and improve the existing work. That is not the spirit of blockchain practitioners as our founder vision, Vitalik Buterin and Gavin Wood. To keep the founder's vision, iExploTech has shared the smart contract and Android Mobile DApps as open-source codes. iExploTech expects other to replicate and improve the work as well as modifying it to meet their organization requirement and practice. By doing so, iExploTech encourages others to contribute what they have found to the public by open source license.

## Conclusion

Blockchain technology is very cheap and convenient to be implemented by any university if they have sufficient technical knowledge and experience rather than just good at writing high-impact publication papers. The main purpose of the IPTM collaboration is to tighten the gap between academia and industry practice. Through the IPTM collaboration, a lecturer and researcher in the university will be able to adapt blockchain sooner and be able to groom graduate students with well-equipped skills that are demanded by the industries.